\documentclass[aps,prd,twocolumn,superscriptaddress,showpacs,amsmath,amssymb]{revtex4}
\usepackage{graphicx,epsf}
\usepackage[]{latexsym}
\newcommand{\be}{\begin{eqnarray}}
\newcommand{\ee}{\end{eqnarray}}

\renewcommand{\d}{\mbox{{\rm d}}}
\begin{document}
%
%
%
\title{Is the Equivalence Principle violated by Generalized Uncertainty
Principles and Holography in a brane-world?}
\author{Fabio~Scardigli}
\email{fabio@yukawa.kyoto-u.ac.jp}
\affiliation{Yukawa Institute for Theoretical Physics, Kyoto
University, Kyoto 606-8502, Japan}
\author{Roberto~Casadio}
\email{casadio@bo.infn.it}
\affiliation{Dipartimento di Fisica, Universit\`a di
Bologna and I.N.F.N., Sezione di Bologna,
via Irnerio~46, 40126 Bologna, Italy}
\begin{abstract}
It has been recently debated whether a class of generalized
uncertainty principles that include gravitational sources of error
are compatible with the holographic principle in models with extra
spatial dimensions.
We had in fact shown elsewhere that the holographic
scaling is lost when more than four space-time dimensions are present.
However, we shall show here that the validity of the holographic
counting can be maintained also in models with extra spatial dimensions,
but at the intriguing price that the equivalence principle for a
point-like source be violated and the inertial mass differ from the
gravitational mass in a specific non-trivial way.
\end{abstract}
\maketitle
%
%
\section{Introduction}
The topic of generalized uncertainty principles (GUP) is a rather
old one. Recently, it has been revived with the addition of gravitational
contributions which provide a minimum length of the order of the
Planck scale (for a review and examples, see~Ref.~\cite{review}). An
attempt in this direction was taken by Ng and van~Dam~\cite{ngvd}
who suggested to include an error due to the space-time curvature
induced by the measuring device, the latter being described, along
the lines of Wigner's 1958 paper~\cite{Wig}, as a system made of a
clock, a photon detector and a photon gun, with total mass $m$ and diameter
$d=2\,a$ (spherical symmetry is assumed for simplicity).
A given length $l$ is then measured by timing the photon travel
from the gun to a suitably placed (ideally weightless) mirror and
back. Photons are also supposed to be emitted in spherical waves,
in order to avoid recoil and back-reaction effects on the clock's position.
This leads to a GUP which yields the remarkable consequence of suggesting that
four-dimensional space-time actually contains (gravitational)
degrees of freedom which scale in agreement with the holographic principle~\cite{holo}.
\par
However, if one tries to extend this result to models with extra spatial
dimensions~\cite{ADD,RS}, the latter property becomes questionable.
It was in fact shown in Ref.~\cite{SC} that a straightforward extension
does not work.
Afterward, different authors suggested to modify the Ng and van~Dam's GUP
by including yet another possible source of error~\cite{Midodashvili}
or by using a black hole for the measuring device~\cite{Maziashvili}.
Both attempts seem to recover the holographic counting.
However, they also require the detector to satisfy rather specific
(indeed peculiar) properties: either it needs to be a black hole, or its
size needs to scale in a very specific way with the mass (see
Ref.~\cite{reply} for more details).
\par
Before we proceed, let us recall why it
is sensible to place on the same footing a ``fundamental'' uncertainty
principle such as Heisenberg's and an (apparently) phenomenological
gravitational source of error.
On general grounds, one understands that in Einstein's general relativity
space-time is a dynamical concept and its quantum description must involve
uncertainty.
Constructions such as that of Ref.~\cite{ngvd} make it clear
that the two sources of uncertainty are closely related:
the photon shot by the gun moves in a Schwarzschild metric with
ADM mass equal to $m$ minus the photon energy $E$~\footnote{This is already
a great simplification of the actual setup, but unfortunately no exact solution
of the Einstein equations is known which describes this two-body problem.}.
Since we are timing the photon's travel,
the time-energy uncertainty relation \footnote{Let us note that we are using the
time-energy uncertainty relation in the standard way, which is usually employed
in order to estimate the life-time of meta-stable states or the width of spectral
lines (see, {\em e.g.}, Ref.~\cite{landau}).}
implies that $E$ has an uncertainty
$\Delta E\sim {\hbar}/{\Delta t_{\rm em}}$ if
$\Delta t_{\rm em}$ is the uncertainty in the time of emission.
Correspondingly, we cannot determine with infinite accuracy
the length of the photon optical path,
say from $r_0>r_{\rm g}$ to $r>r_0$
(in the detector's frame), with $r_{\rm g}(m)={2\,G_{\rm N}\, m}/{c^2}$,
but can just find the lower and upper bounds
\be
\int\limits_{r_0}^{r}
\!\!
\frac{\d \rho}{1-\frac{r_{\rm g}^+}{\rho}}
\equiv
c\,\Delta t_{\rm max}
\gtrsim
c\,\Delta t
\gtrsim
c\,\Delta t_{\rm min}
\equiv
\int\limits_{r_0}^{r}
\!\!
\frac{\d \rho}{1-\frac{r_{\rm g}^-}{\rho}}
\ ,
\ee
where $r_{\rm g}^\pm=r_{\rm g}(m-E\pm\Delta E)$.
Ref.~\cite{ngvd} then suggests to add to other sources of errors
the uncertainty in the length of the optical path as the difference
\be
\delta l_{\rm C}\simeq
c\,(\Delta t_{\rm max}-\Delta t_{\rm min})
\ .
\label{lC}
\ee
\par
The aim of this paper is to take the opposite perspective with
respect to some previous works and {\em to show\/} that the GUP of
Refs.~\cite{ngvd,SC} and the holographic principle {\em can\/} be both
kept valid consistently.
However, we shall then show that this leads to another principle being
violated, namely the detector's inertial mass and gravitational mass
must differ in models with extra spatial dimensions.
We shall write explicitly the fundamental constants $c$, $\hbar$
and Newton's constant $G_{\rm N}$ or, alternatively, the Planck length
$\ell_{\rm p}=({G_{\rm N}\,\hbar}/{c^3})^{1/2}$
or mass $M_{\rm p}=\hbar/2\,c\,\ell_{\rm p}$
[respectively replaced by
$G_{(n+4)}$, $\ell_{(4+n)}=({G_{(4+n)}\,\hbar}/{c^3})^{\frac{1}{2+n}}$
and $M_{(4+n)}=\hbar/2\,c\,\ell_{(4+n)}$ in $4+n$ dimensions].
%
%
%
\section{Gravitational GUP}
Suppose we wish to measure a distance $l$ with the detector
described in the {\em Introduction}.
If $\Delta x$ is the initial uncertainty in the
position of the clock, after the time $T = 2\,l/c$
taken by the photon to return to the detector, the uncertainty in
the actual length of the segment $l$ becomes
\be
\Delta x_{\rm tot}=\Delta x+T\,\Delta v
=\Delta x+\frac{\hbar\,T}{2\,m\,\Delta x}
\ ,
\ee
where $\Delta v$ is the uncertainty in the detector's
velocity from Heisenberg's principle.
Upon minimizing $\Delta x_{\rm tot}$ with respect to $\Delta x$
we obtain Wigner's quantum mechanical error~\cite{Wig}
\be
\delta l_{\rm QM}\simeq
2\left(\frac{\hbar\,l}{m\,c}\right)^{1/2}
\ ,
\label{dl_QM}
\ee
which we seem to be able to reduce
as much as we want by choosing $m$ very large.
But gravity now gets in the way as mentioned before.
\par
An important remark is that the measuring device cannot be a black hole,
or it could not serve as a photon gun~\cite{AC},
\be
a>r_{\rm g}
\quad
\Rightarrow
\quad
\delta l_{\rm QM}^2
\gtrsim
8\,\ell_{\rm p}^2\,\frac{l}{a}
\label{eqm}
\ .
\ee
Besides, we need now to include in the computation the gravitational
error of Eq.~(\ref{lC}) with
$r=a+l>r_0=a\gg r_{\rm g}$.
We consider for $r_{\rm g}^\pm$ the lower and upper bounds
allowed by total energy conservation, corresponding to the two
limiting cases $E-\Delta E=0$ and $E+\Delta E=m$, respectively.
This yields, for distances $l\gtrsim a$,
\be
\delta l_{\rm C}\simeq
r_{\rm g}\,\log\left(\frac{a+l}{a}\right)
\gtrsim r_{\rm g} \log 2 \simeq \frac{r_{\rm g}}{2}
\, .
\ee
Note that $\delta l_{\rm C}$ increases with increasing detector's mass
and the total error becomes
\be
\delta l_{\rm tot} = \delta l_{\rm QM} + \delta l_{\rm C}
\simeq
2\left(\frac{\hbar\,l}{m\,c}\right)^{1/2} + \frac{G_{\rm N}\,m}{c^2}
\ .
\label{dl_tot4}
\ee
For a given $l$, this error can only be minimized with respect to
the mass of the clock, which yields
\be
\left(\delta l_{\rm tot}\right)_{\rm min}
\simeq
3\left(\ell_{\rm p}^2\,l\right)^{1/3}
\ ,
\label{otto}
\ee
for $m=2\,M_{\rm p}(l/l_{\rm p})^{1/3}$.
The global uncertainty on $l$ therefore contains precisely the term
proportional to $l^{1/3}$ required by the holography~\footnote{It easy
to check that the quantum mechanical bound in Eq.~(\ref{eqm})
is always smaller than the global uncertainty~(\ref{otto}), as long as the clock's
size $a$ is larger than~(\ref{otto}) itself.
In fact, $2\,\sqrt{2}\,\ell_{\rm p}\,(l/a)^{1/2}<3\,(\ell_{\rm p}^2 \,l)^{1/3}$ iff
$a > (8/9)(\ell_{\rm p}^2 \,l)^{1/3}$, a condition easily met in all meaningful
cases.}.
\par
Unfortunately, in $4+n$ dimensions this does not seem to work.
When $a+l$ is shorter than the size $L$ of the extra dimensions,
one can use the $4+n$-dimensional Schwarzschild metric~\cite{MP}
\be
\d s^2
&\!=\!&
g_{\mu\nu}\,\d x^\mu\,\d x^\nu
\nonumber
\\
&\!=\!&
-F(r)\,c^2\,\d t^2+F(r)^{-1}\,\d r^2+r^2\,\d\Omega_{n+2}^2
\ ,
\label{S4pd}
\ee
where Greek indices run from 0 to $3+n$ (Latin indices
will denote spatial coordinates) with
\begin{subequations}
\be
F(r)=1-{C}/{r^{1+n}}
\ ,
\ee
and
\be
C=\frac{16\,\pi\,G_{4+n}\,m}{(2+n)\,A_{2+n}\,c^2}
\ ,
\label{C}
\ee
\end{subequations}
$A_{2+n}$ being the surface area of the unit $(2+n)$-sphere.
Upon repeating analogous steps, one then finds~\cite{SC}
\be
\left(\delta l_{\rm tot}\right)_{\rm min} \sim
\left(a^{-n}\,{\ell_{(4+n)}^{2+n}\,l}\right)^{1/3}
\ .
\label{dl1}
\ee
The above expression, even in the rather ideal case $a \sim \ell_{(4+n)}$,
yields the following scaling for the number of degrees of freedom
in a volume $V$ of size $l$,
\be
\mathcal{N}(V)
=
\left(\frac{l}{\left(\delta l_{\rm tot}\right)_{\rm min}}\right)^{3+n}
\sim
\left(\frac{l}{\ell_{(4+n)}}\right)^{2\,\left(1+\frac{n}{3}\right)}
\ ,
\label{N}
\ee
and the holographic counting holds in four-dimensions ($n=0$) but is lost
when $n>0$.
%
%
%
%
\section{GUP, Holography and the Equivalence Principle}
Let us now point out that, beside the GUP proposed by Ng and van~Dam,
the result in Eq.~(\ref{dl1}) deeply relies on the use of the black hole
metric~(\ref{S4pd}) and its dependence on the mass $m$.
In particular, the expression for the parameter $C$ is obtained by taking the
weak field limit~\cite{MP} in which the metric can be written as
$g_{\mu\nu}=\eta_{\mu\nu}+h_{\mu\nu}$, with $|h_{\mu\nu}|\ll 1$ in the
asymptotic region far from any source.
The linearized metric $h_{\mu\nu}$, in the harmonic gauge,
obeys the Poisson equation
\be
\nabla^2 h_{\mu\nu}=-\frac{16\,\pi\,G_{4+n}}{c^4}\,\bar{T}_{\mu\nu}
\ ,
\label{hg}
\ee
with a source $\bar{T}_{\mu\nu}$ related to the stress-energy
tensor by
\be
\bar{T}_{\mu\nu}=\left(T_{\mu\nu}-\frac{1}{2+n}\,\eta_{\mu\nu}\,T\right)
\ .
\ee
The condition that the system be non-relativistic means that time
derivatives can be considered much smaller than spatial derivatives,
so that the components of the stress energy tensor can
be ordered as $|T_{00}|\gg|T_{0i}|\gg|T_{ij}|$.
A solution to Eq.~(\ref{hg}) is then given by
\be
\!\!\!\!\!\!\!\!
h_{\mu\nu}(x)&\!=\!&
\frac{16\,\pi\,G_{4+n}}{(1+n)\,A_{2+n}\,c^4}\,
\int\frac{\bar{T}_{\mu\nu}(y)}{|x-y|^{1+n}}\,
\d^{3+n} y
\nonumber
\\
&\!\simeq\! &
\frac{16\,\pi\,G_{4+n}}{(1+n)\,A_{2+n}\,c^4}\frac{1}{r^{1+n}}
\int\bar{T}_{\mu\nu}\,\d^{3+n} y
\nonumber
\\
&&
+\frac{16\,\pi\,G_{4+n}}{A_{2+n}\,c^4}\,\frac{x^k}{r^{3+n}}
\int y^k\,\bar{T}_{\mu\nu}\,\d^{3+n} y
+\dots
\ ,
\ee
where the approximate equality is obtained by expanding for $r=|x|\gg |y|$.
Myers and Perry define the $4+n$-dimensional ADM mass $m$ as
\be
\int\bar{T}_{00}\,\d^{3+n}x = m\,c^2
\ ,
\ee
so that one obtains the natural generalization of the Newtonian potential
to $4+n$ dimensions,
\be
h_{00}\simeq
\frac{16\,\pi\,G_{4+n}}{(2+n)\,A_{2+n}\,c^2}\,\frac{m}{r^{1+n}}
=\frac{C}{r^{1+n}}
\ .
\ee
\par
One can now wonder if the metric defined by Eqs.~(\ref{S4pd})-(\ref{C})
can be modified in such a way that the holographic principle be fulfilled also in
$4+n$ dimensions, thus suitably changing the counting of degrees of freedom
given in Eq.~(\ref{N}).
In other words, we shall assume the holographic principle as a constraint to fix
the form of the $4+n$-dimensional black hole metric.
Of course, this new metric must still satisfy the $4+n$-dimensional Einstein
equations~(\ref{hg}), which is a very strong constraint and
it seems therefore sensible to change the original metric~(\ref{S4pd})-(\ref{C})
as little as possible.
On the other hand, we note that the Myers-Perry solution exhibits a complete
$3+n$-dimensional spherical symmetry, which means that it ignores the weight
of the brane~\footnote{The effect of the brane on black holes has never been
computed exactly.
There are however both numerical~\cite{num} and analytical~\cite{CM}
works which show that it should squeeze the horizon along the extra
dimension(s), thus breaking the full spherical symmetry.}.
All things considered, the deformation of the metric~(\ref{S4pd})-(\ref{C})
which we shall use consists in allowing for a departure from a linear relation
between the inertial mass and the gravitational ADM mass of the form
\be
\int\bar{T}_{00}\,\d^{3+n} x =
M_{(4+n)}\,c^2\,\left(\frac{m}{M_{(4+n)}}\right)^{\gamma(n)}
\ ,
\label{M}
\ee
where $\gamma$ is a (yet) unspecified function of $n$.
Although this {\em ansatz\/} is not the only one that can
in principle be conceived, it really is one of the simplest possible,
as Eq.~(\ref{M}) means that the gravitational mass $M$ and inertial mass
$m$ of the source (the detector) are related by
\be
M=M_{(4+n)}\,\left(\frac{m}{M_{(4+n)}}\right)^{\gamma(n)}
\ .
\label{mg}
\ee
The equivalence principle would thus be violated for any function
$\gamma\not=1$.
\par
Eq.~\eqref{mg}  yields a total error in length measurements
given by
\be
\delta l_{\rm tot} = \delta l_{\rm QM} + \delta l_{\rm C} \simeq
\frac{J}{\sqrt{m}} + K m^\gamma
\ ,
\label{dl_tot}
\ee
where
\be
J=2\left(\frac{\hbar\,l}{c}\right)^{1/2}
\!\! ,
\quad
K=\frac{2^n-1}{n\,2^n\,a^n}\,
\frac{16\,\pi\,G_{4+n}\,M_{4+n}^{(1-\gamma)}}{(2+n)\,A_{2+n}\,c^2}
\ .
\ee
Upon minimizing $\delta l_{\rm tot}$ with respect to $m$, one obtains
\be
\left(\delta l_{\rm tot}\right)_{\rm min}
\sim
l^{\frac{\gamma}{2\,\gamma+1}}
\ .
\ee
Hence, if we require that holography holds, namely
$\left(\delta l_{\rm tot}\right)_{\rm min}\sim
\left( l\right)^{\frac{1}{3+n}}$, we must also have
\be
\frac{\gamma}{2\,\gamma+1} = \frac{1}{3+n}
\ .
\ee
In this way we see that the holographic scaling can be preserved
also in $4+n$ dimensions, with a Schwarzschild-like metric for point-like
sources, provided we define the gravitational mass $M$
as in Eq.~(\ref{mg}) with~\footnote{Note that gravity does not propagate
in less than four dimensions and our results are therefore
expected to hold only for $n\ge 0$~\cite{SC} (for which $\gamma>0$ and
$M$ is finite).}
\be
\gamma = \frac{1}{1+n}
\ .
\ee
Therefore, the equivalence principle must be violated
at distances shorter than the size $L$ of the extra dimensions,
as well as Newton's law is modified in $4+n$ dimensions
({\em i.e.}, $F\sim 1/r^{2+n}$).
%
%
%
%
\section{Yet another view}
For the sake of completeness and in order to further support our results,
we report and comment hereafter on a more recent proposal of Ng's.
In Ref.~\cite{ng2}, he describes a different argument to reconcile
GUP with holography.
He says:
``...To see this, let the clock be a
light-clock consisting of a spherical cavity of diameter $d=2\,a$,
surrounded by a mirror wall of mass $m$, between which bounces a
beam of light.
For the uncertainty in distance measurement not to be greater than
$\delta l$, the clock must tick off time fast enough that
$d/c \lesssim \delta l/c$.
But $d$, the size of the clock, must be larger than the
Schwarzschild radius $r_S \equiv 2\,G_{\rm N}\,m/c^2$ of the mirror,
for otherwise one cannot read the time registered on the clock.
From these two requirements, it follows that
\be
\delta l \gtrsim {G_{\rm N}\,m}/{c^2}
\ .
\ee
Thus general relativity alone would suggest using a light clock to
do the measurement.''
\par
On combining the last inequality with Wigner's bound in Eq.~(\ref{dl_QM}),
Ng readily obtains the expression in Eq.~(\ref{dl_tot4}) and then,
minimizing with respect to $m$, the result~(\ref{otto}) for the
minimum total error.
This works flawlessly in four dimensions, and, again in Ref.~\cite{ng2},
Ng applies the same kind of argument also to space-times with
$4+n$ dimensions.
He considers the clock as completely immersed in the extra dimensions,
that is $0<d<L$, so that the metric~(\ref{S4pd}) can be used.
The condition that the clock be not a black hole,
\be
a > r_{\rm g}=C^{\frac{1}{1+n}}
\ ,
\label{bh}
\ee
and the fact that the error must be larger than the size
of the clock, $\delta l \gtrsim d$, then imply that
\be
\delta l \gtrsim C^{\frac{1}{1+n}}
\ .
\label{wrong}
\ee
This means (including Wigner's quantum error and omitting unimportant
numerical factors) that
\be
\delta l \gtrsim \left(\frac{\hbar\,l}{m\,c}\right)^{1/2} + C^{\frac{1}{1+n}}
\ .
\label{dl}
\ee
Since, from Eq.~(\ref{C}), $C\sim m$, after extremizing for $m$, Ng
obtains the total error
\be
\delta l_{\rm tot} \sim
\left(\ell_{(4+n)}^{2+n}\,l\right)^{\frac{1}{3+n}}
\ ,
\ee
so that the holographic scaling is satisfied also in $4+n$ dimensions.
Of course, this result is in sharp contrast with the one obtained in
Ref.~\cite{SC}.
\par
The argument of Ng can be however criticized in the following way.
The key point is again, like in Ref.~\cite{Midodashvili},
the relation between the size of the clock $d$ and its mass $m$.
As we have shown in Ref.~\cite{reply}, the size of the clock can be considered as
an error
(more precisely, the actual error $b$ is very likely much smaller than $d$,
{\em i.e.}, $b \ll d$),
provided one also considers that, to all practical purposes, there is no
universal relation between the size $d$ of a clock and its mass $m$.
Therefore, although the inequality~(\ref{bh}) should in general hold
(otherwise the clock would be a black hole),
it cannot be used to establish~(\ref{wrong}) as a {\em universal\/}
expression of the error $\delta l$ in terms of the clock's mass $m$.
In fact, on following a similar logic, one could then consider as a valid
lower bound for the error $\delta l$ any expression containing the mass
of the clock itself, provided it limits from below the size of the clock.
For example, one could require that
the size $d$ be larger than the clock's Compton wavelength, $d>\hbar/m\,c$,
in order to have a classical clock.
This would immediately yield a completely different (and {\em non-holographic\/})
scaling of the total error $\delta l_{\rm tot}$.
Therefore, the use of the Schwarzschild radius of the clock in the expression~(\ref{dl})
as a measure of the gravitational part of the error appears to be completely arbitrary.
On the other hand, if one correctly considers the size of the clock as a contribution
to the error, but strictly
{\em independent of\/} $m$,
\be
\delta l \gtrsim \left(\frac{\hbar\,l}{m\,c}\right)^{1/2} + d
\ ,
\ee
then the holographic scaling in $4+n$ dimensions is not recovered~\cite{reply}.
\par
Finally, we point out that, even adopting a device like that of
Refs.~\cite{ng2}, a gravitational error originated by the uncertainty
in the ADM mass of the Schwarzschild metric of the kind discussed in
the {\em Introduction\/} should still be included.
The complete and correct final expression of the total error would therefore be
\be
\delta l_{\rm tot} = \delta l_{\rm QM} + \delta l_{\rm C}
\simeq
\frac{J}{\sqrt{m}} + K\, m^\gamma + b
\ ,
\ee
where $b$ ($\ll d$) does not depend on $m$.
This would again yield the same conclusion following from
Eq.~(\ref{dl_tot}).
%
%
%
%
\section{Conclusions}
We have shown how a gravitational error originated by the quantum mechanical
uncertainty in the ADM mass of a detector inevitably affects any measurements
of length.
This leads to Ng and van Dam's GUP, which has the remarkable property
of respecting the holographic counting in four dimensions.
When extra spatial dimensions are present, the holographic scaling is
violated.
However, holography can be restored if one instead allows for
a violation of the equivalence principle at short distances (below the
size of extra dimensions).
This violation could in principle be tested (see, {\em e.g.}, Ref.~\cite{lamm}),
and its extent is related to the number of extra dimensions.
The connections of the present scenario with other models where the equivalence
principle is also violated are worth of further investigation.
To this aim, the results reported for example in Refs.~\cite{vari} seem to be
particularly promising.
Such results, although sometimes worked out in a stringy oriented scenario
(for example D-brane induced gravity) or in the framework of loop quantum gravity,
seem anyway to match, at least for the key aspects, the more phenomenological
arguments given here.
\par
%
\acknowledgements
F.~S.~gratefully thanks C.~Germani for many enlightening discussions on the subject
and D.A.M.T.P., Cambridge UK, for hospitality during the early stages of this work.
Fruitful discussions with R.~Adler, P.~Chen and S.~Carlip are also acknowledged.
\end{document}